\documentclass[12pt,nofootinbib]{iopart}

\newcommand{\mrm}[1]{_{\rm #1}}
\renewcommand{\d}{{\rm d}}
\usepackage{iopams}
\usepackage{graphicx}
\usepackage{url}
\usepackage{envmath}
\usepackage{aas_macros}
\usepackage{setstack}
\usepackage{color}

\begin{document}

\title[Detecting Planet 9 via Hawking radiation]{Detecting Planet 9 via Hawking radiation}

\author{Alexandre Arbey$^{1,2}$ and J\'er\'emy Auffinger$^1$}

\address{$^1$ Univ Lyon, Univ Claude Bernard Lyon 1, CNRS/IN2P3, IP2I Lyon, UMR 5822, F-69622, Villeurbanne, France}
\address{$^2$ Institut Universitaire de France (IUF), 103 boulevard Saint-Michel, 75005 Paris, France}
\ead{j.auffinger@ipnl.in2p3.fr}

\begin{abstract}
Concordant evidence points towards the existence of a ninth planet in the Solar System at more than $400\,$AU from the Sun. In particular, trans-Neptunian object orbits are perturbed by the presence of a putative gravitational source.
Since this planet has not yet been observationally found with conventional telescope research, it has been argued that it could be a dark compact object, namely a black hole of probably primordial origin.
Within this assumption, we discuss the possibility of detecting Planet 9 via a sub-relativistic spacecraft fly-by and the measure of its Hawking radiation in the radio band and conclude that it is too faint compared to the CMB. We thus present other perspectives with rather a satellite mission and conclude that smaller black holes would give much more interesting signals. We emphasize the importance of the study of such Hawking radiation laboratories in the Solar System.
\end{abstract}
\vspace{2pc}

\section{Introduction}

Perturbations of orbits of known objects in the Solar System have led astronomers to search for gravitational sources from which they originate, under the form of unknown planets. After the discovery of Neptune in 1846, no more planets were found beyond dwarf planets such as Pluto or Eris. However concordant evidence have recently appeared in direction of what has been a proofless obsession for many astronomers: the existence of Planet 9, which may become an object under even more intense scrutiny. The apparent clustering of trans-Neptunian objects (TNOs) orbits in the Kuiper belt suggests the presence of a massive body of a mass $M\sim 5-10\, M_\oplus$ orbiting between $300$ and $1000\,$AU \cite{Planet9_2016,Planet9_2019}. Even though the statistics of clustered TNOs is not sufficient enough to robustly exclude coincidental observations, the probability of accidental correlations is $\lesssim 1\%$ \cite{Clement2020}. The parameters of this hypothetical Planet 9 are further constrained by ephemeride measurements such as those of Cassini \cite{Cassini2016,Cassini2016_2}.

In spite of telescope searches, no new object has been found in the sky to be Planet~9. Ref.~\cite{Scholtz2019} thus suggests that Planet 9 may be a compact dark object, invisible to telescopes -- namely, a Black Hole (BH). A BH with such a light mass certainly points towards a non-stellar origin because of the Tolman-Oppenheimer-Volkoff limit; this BH could be one of the putative primordial BHs (PBHs) that are under intense scrutiny since they could represent some or all of dark matter (DM) (for a recent review on PBH formation mechanisms and constraints, see e.g. \cite{Carr2020} and references therein). PBH abundance is severely constrained for about 50 orders of magnitude in mass, but there still exists an open parameter space for them to represent all DM in the sub-lunar mass range, or part of it in various other mass windows. The fraction of dark matter under the form of PBHs is expected to be $f\sim 0.1-0.01$ in the Planet 9 mass region \cite{Carr2020}. PBHs are believed to have formed after inflation from primordial density inhomogeneities that collapsed when the overdensity was above some threshold. No confirmed PBH has been observed yet, but OGLE has recently found PBH candidates in microlensing events \cite{OGLE2017} whose masses would correspond to the mass of Planet 9, and their mergers may be detectable in future gravitational wave experiments \cite{Miller2020,Herman2020}. Thus, it is plausible to consider that if a population of terrestrial mass PBHs exists, one of them could have been captured by the Sun gravity and could be orbiting beyond Neptune, providing an explanation for the ``invisible" body responsible for the gravitational anomalies of TNOs.

Successively, two experiments have been proposed to detect Planet 9 if it were a BH (hereafter called P9). Both are based on ideas similar to the Breakthrough Starshot proposal\footnote{\url{https://breakthroughinitiatives.org/Initiative/3}}, in which it is proposed to send a fleet of very small spacecrafts ($m\sim\,$g$-$kg) at sub-relativistic speeds ($v\sim 0.001c$) in different directions of the sky to reach nearby stars in order to study their planetary systems and achieve the most distant explorations ever \cite{Starshot2020}. Their advantage is that such light and fast spacecrafts would reach the orbit of an eventual P9 in a few years. By sending many of those across the sky towards the hypothetical location of P9 orbit, one gets a chance that one of them experiences a fly-by of P9. The first proposal is to measure the time delay in the line of sight trajectory of a given spacecraft (hereafter called SC0 for spacecraft 0, the discoverer), induced by the presence of a nearby massive body \cite{P9_2020_1}. This would necessitate an on-board precision clock to measure a $\sim 10^{-5}\,$s time delay over a one year trajectory. The second proposal is to measure the transverse inclination of the trajectory of SC0 induced by the presence of P9 \cite{P9_2020_2}. This alleviates the on-board clock problem but necessitates a $\sim 10^{-9}\,$rad angular displacement measurement, which could be achievable with VLBI for example. However, in Ref.~\cite{P9_2020_3} the authors examined the environment in which SC0 would travel to reach the orbit of P9 and concluded that the interstellar medium turbulence -- drag and magnetic fields -- would make the precise gravitation-perturbed trajectory measurements cited above impossible to achieve due to noise signals from unknown medium local properties.

There also exists a completely different approach to P9 detection proposed in \cite{P9_2020_4}, based on the fact that icy objects of the Oort cloud would get disrupted by the P9 gravitational field and the accretion of such material could cause flares detectable by the LSST survey\footnote{\url{https://www.lsst.org/}} \cite{LSST2019}. A few of such events could occur per year, making them detectable. In addition, it would prove the BH nature of P9, and solve the trajectory difficulties of the sub-relativistic spacecrafts described in \cite{P9_2020_3}. The continuous search for TNOs also continues, and the DES collaboration claims that they would be able to detect many more of them, among which more clustered TNOs pleading in favor of P9, if not P9 itself \cite{Bernardinelli2020,henghes2021}.

Here we suggest a new proposal, based on the fact that P9, if it is indeed a BH, will emit Hawking radiation \cite{Hawking1975}. When classical general relativity is mixed up with quantum mechanics effects, the fluctuations of the vacuum at the horizon of a BH give rise to a net emission of particles at spatial infinity, causing the BH to slowly evaporate away. Thus even if P9 is not visible from the Earth (not being a reflective planet but a BH), it would still emit a small amount of radiation. This was already considered in the original paper about the BH nature of P9 \cite{Scholtz2019} but the authors concluded that the amount of Hawking radiation was too small to be detectable \textit{from Earth}, which is true. What we consider here is the detection of this very Hawking radiation by the flying-by SC0, \textit{in the vicinity} of P9, as described in the next section. This would be of particular importance since, even if rather well theoretically motivated, Hawking radiation has not yet been observed, because the power received on Earth is much too small for conventional BHs, such as stellar ones like Cygnus X-1 \cite{Cygnus_1,Cygnus_2,Cygnus_3} or supermassive ones like Sagitarius A$^*$ \cite{SagitariusA2019,SagitariusA2020}. Hawking radiation by smaller BHs results in constraints on their abundance but not in detection signals, see e.g.$\!\!$ the recent work on BBN \cite{Carr2020,Constraints_BBN}, CMB \cite{Constraints_BBN,Constraints_CMB,Constraints_CMB_2}, gamma rays \cite{Constraints_gamma_1,Constraints_gamma_2,Constraints_gamma_3}, electrons-positrons annihilation or detection \cite{Constraints_electrons,Constraints_electrons_1,Constraints_electrons_2,Constraints_neutrinos_2}, neutrinos \cite{Constraints_neutrinos_2,Constraints_neutrinos_1}, local PBH burst rate \cite{Constraints_burst_1,Constraints_burst_2,Constraints_burst_3}, dark matter production \cite{Constraints_DM_2,Constraints_DM} and even primordial gravitational waves \cite{GW2020,GW2020_2}. Nevertheless the precise spectrum of Hawking radiation may contain information on the quantum structure of BH horizons. Therefore directly observing the BH Hawking radiation would be of great importance, and a PBH in our Solar System would represent the best laboratory to study it.

In the first section of this paper, we will discuss the detectability of P9 during a fly-by based on its electromagnetic Hawking radiation, and in the second section we will present other perspectives like satellization of a spacecraft or search for smaller PBHs in the solar system.

\section{Hawking radiation light curves}

\subsection{Setup}

The setup of the experiment would be the following. SC0 passes by P9 at speed $v$ and with impact parameter $b$. We define $t=0$ to be the time of minimal approach. When SC0 approaches P9, the radiation flux will increase, reach maximum at $t=0$ and then decrease. Since we consider sub-relativistic velocities, Doppler effect is negligible. The spatial displacements considered in \cite{P9_2020_1,P9_2020_2,P9_2020_3} have however to be taken into account as an uncertainty on the precise trajectory of the ship. We neglect them for the moment and consider an ideal straight line trajectory for SC0.

\begin{figure}[t]
	\centering
	\includegraphics[scale = 1]{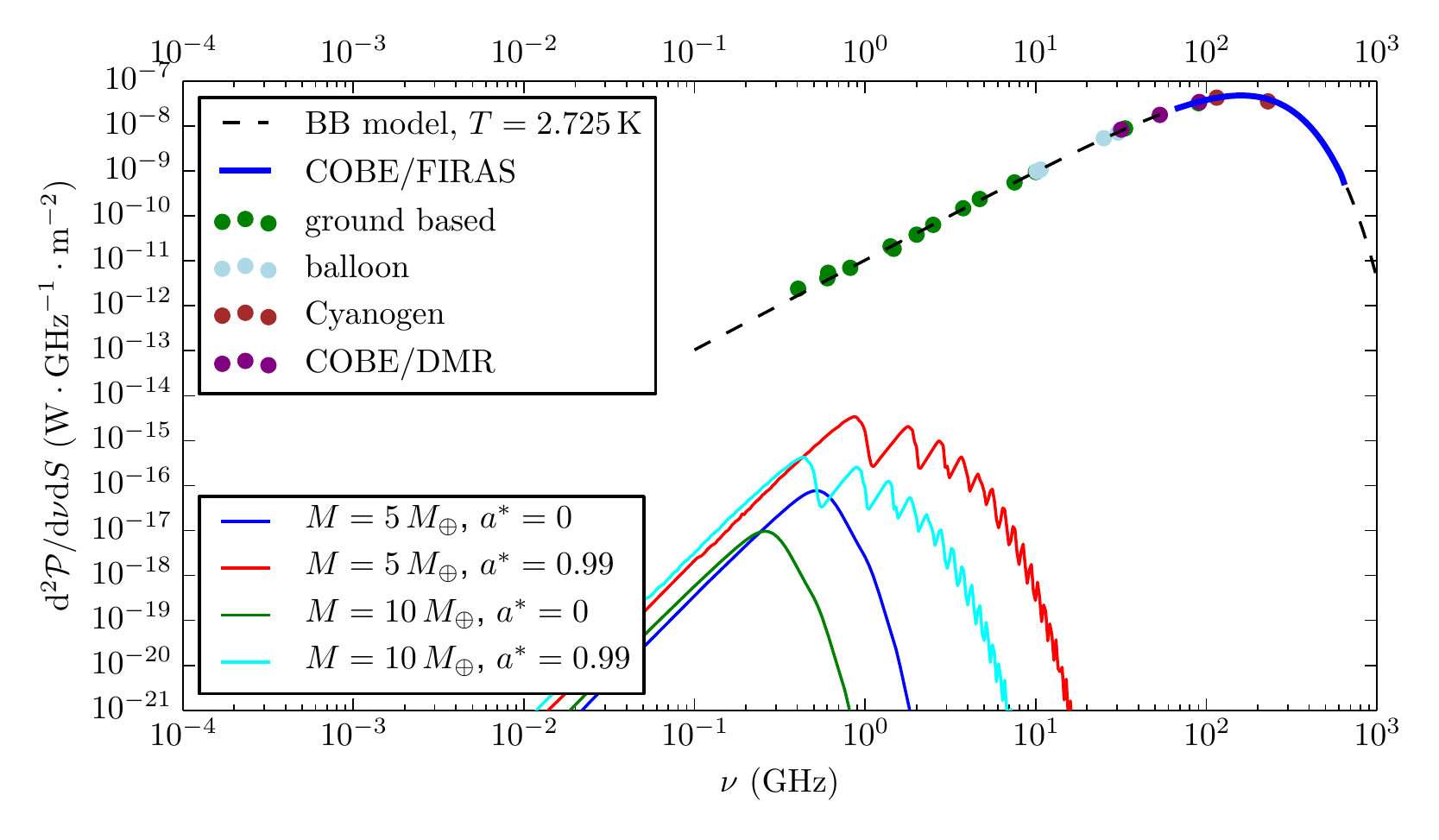}
	\caption{Total power emission of photons by P9 as a function of frequency for different values of the P9 parameters $M = \{5,10\}M_\oplus$ and $a^* = \{0,0.99\}$ (bottom solid lines). The CMB spectrum is also shown for comparison, as measured by different instruments \cite{FIRAS} (dots and thick blue line) and with a blackbody model fit with temperature $T = 2.725\,$K (dashed black line).}
	\label{fig:band}
\end{figure}

P9 has a super-terrestrial mass $M\mrm{P9} \sim 5-10\,M_\oplus$, thus its peak electromagnetic emission frequency lies around the GHz radio band. We do not have any indication of P9 dimensionless spin $a^*$; as a PBH it is expected to have a negligible spin but it has been shown that transient matter-domination era at the end of inflation can produce high-spin PBHs that can conserve their spin until today despite Hawking evaporation \cite{Arbey2020}. We show in Fig.~\ref{fig:band} the power emission per horizon area unit as a function of frequency for different P9 masses and spins such as
\begin{equation}
    \frac{\d^2\mathcal{P}}{\d \nu \d S} = \frac{1}{4\pi r\mrm{S}^2}E\frac{\d^2 N}{\d t\d \nu}\,,
\end{equation}
where ${\d^2 N}/{\d t\d \nu}$ is the number of photons emitted by Hawking radiation per units of time and frequency and $r\mrm{S} = 2M$. We clearly see that the low-mass high-spin setup is favored by detection because it implies more energetic and abundant emission. In this figure we also show the CMB spectrum which is very well approximated by a blackbody radiation
\begin{equation}
    \frac{\d^2\mathcal{ P}\mrm{CMB}}{\d \nu \d S} = \frac{8\pi\nu^3}{e^{E/T\mrm{CMB}} - 1}\,,
\end{equation}
with temperature $T\mrm{CMB} = 2.725\,$K \cite{Planck2018}. The comparison with the CMB intensity in the Hawking radiation energy range show that even in the most favorable case, P9 radiation represents $\sim 0.1-1\%$ of the cosmic microwave background (CMB) intensity, a difficulty that will be discussed below. Indeed, a PBH with mass $M\sim 5\,M_\oplus$ has a temperature of $T\sim 2\times 10^{-3}\,T\mrm{CMB}$, therefore its Hawking radiation is subdominant compared to the CMB. The wiggles appearing for the Hawking radiation of high spin BHs are not artifacts of the numerical computation, they are due to the quantized coupling between the BH spin and the emitted radiation angular momentum \cite{Dong2016} and get significant once one reaches quasi extremal spins with $a^* \gtrsim 0.9$.

Let us consider that the solar sail of the Breakthrough Starshot-like spacecrafts considered here is used as a radio antenna in the GHz band, with a surface area of $\mathcal{S} \sim\,$m$^2$ \cite{Starshot2020}. As Breakthrough Starshot already considers very large sails, it is natural to imagine a way to use these sails as on-board detectors to perform electromagnetic measurements. This would require to adapt the sail technology to implement GHz photon collecting, a possible challenge as the sails have to be extremely thin and light for solar propulsion into such small spacecrafts. The power received by the ship, if its sail is considered perpendicular to its trajectory, is then of the form
\begin{equation}
\mathcal{P}(t) = \eta\frac{S(t)}{4\pi r(t)^2} \int_{0}^{+\infty} E \frac{\d^2 N}{\d t\d E} \d E\,, \label{eq:power}
\end{equation}
where the energy integral covers the radio GHz band, $r(t)$ is the distance between SC0 and P9 and $S(t)$ is the area of the sail projected in the direction of P9. Here ${\d^2 N}/{\d t\d E}$ is the number of photons emitted per units of time and energy. The emission rates of particles by evaporating BHs are computed using the public code \texttt{BlackHawk} \cite{BlackHawk2019}. The efficiency coefficient $\eta$ corresponding to the absorption of the sail is considered in Eq.~(\ref{eq:power}) for completeness, but since we do not make any assumption on the material or technology, we do not have an estimation of it; in any case it has to be maximized. Finally we assume the sail to be perpendicular to the direction of motion for simplicity, but we note that there probably exists more optimized geometries to maximize the power received during a fly-by while keeping a sufficient acceleration via laser propulsion.

\subsection{Ideal straight line trajectory}

We geometrically compute $S(t)$ and $r(t)$ by defining $\alpha$ as the angle between SC0 velocity $\mathbf{v}$ and position $\mathbf{r}$ relative to the origin at P9, and consider that the (one dimensional) sides of an area $A$ have lengths of the order $\sqrt{A}$. We obtain 
\begin{equation}
\cos(\alpha) = \frac{\sqrt{S(t)}}{\sqrt{\mathcal{S}}} \iff S(t) = \cos(\alpha)^2 \mathcal{S}\,,
\end{equation}
and
\begin{equation}
\tan(\alpha) = \frac{b - \sqrt{\mathcal{S}}}{|r^*(t)|}\,,
\end{equation}
where the $-\sqrt{\mathcal{S}}$ term comes from the definition of $b$ which is the distance between P9 and the \emph{center} of the solar sail. This factor can be safely neglected. Thus the projected area is
\begin{equation}
S(t) = \cos\left[\arctan\left( \frac{b - \sqrt{\mathcal{S}}}{|r^*(t)|} \right)\right]^2\mathcal{S}\,, \label{eq:area}
\end{equation}
where $r^*(t) = vt$ is the distance to minimal approach in the straight trajectory approximation and $r(t) = \sqrt{r^*(t)^2 + b^2}$. We see that even if the distance is minimal at ($t = 0 , r^*(t) = 0,\, r(t) = b$), the projection of the flux on the sail is zero at this point. Thus we expect a peak feature in the time-dependent radio signal with a discontinuity at $t = 0$.

\subsection{Perturbed trajectory}

If the kinetic energy carried by SC0 becomes comparable to the gravitational potential energy of P9, we can expect a gravitational perturbation of the trajectory, i.e. for
\begin{equation}
E\mrm{kin} \sim E\mrm{pot} \iff \frac{1}{2}mv^2 \sim \frac{GMm}{r} \iff r \sim \frac{2GM}{v^2}\,.
\end{equation}
Considering the speed and mass at stake here, it occurs when $b\underset{\sim}{<} 100\,$km. The trajectory will be deviated as given in \cite{P9_2020_2,P9_2020_3} because of the time build-up of small shifts, but this will occur at timescales much larger than this fly-by detection time. However if the impact parameter becomes very small the full trajectory needs to be taken into account to predict the form of the signal. This can be done by taking again the geometrical definitions given in the previous section and redefining an effective instantaneous (at time $t$) impact parameter $\overline{b}(t)$ and effective instantaneous distance to the minimal approach point $\overline{r}^*(t)$, which could be seen as the geometric quantities obtained in case SC0 were to continue in a straight line from time $t$. Thus the $\overline{\alpha}(t)$ angle is the angle between the instantaneous velocity and position vectors
\begin{equation}
\cos(\overline{\alpha}) = \frac{\mathbf{v}\cdot\mathbf{r}}{vr}\,,
\end{equation}
and the perturbed quantities to be considered in the area projection formula in Eq.~(\ref{eq:area}) are
\begin{equation}
\overline{b} = r\sin(\overline{\alpha}) \,, \qquad\qquad
\overline{r}^* = r\cos(\overline{\alpha}) \,.
\end{equation}

\subsection{Results}

\begin{figure}[t]
	\centering
	\includegraphics[scale = 1]{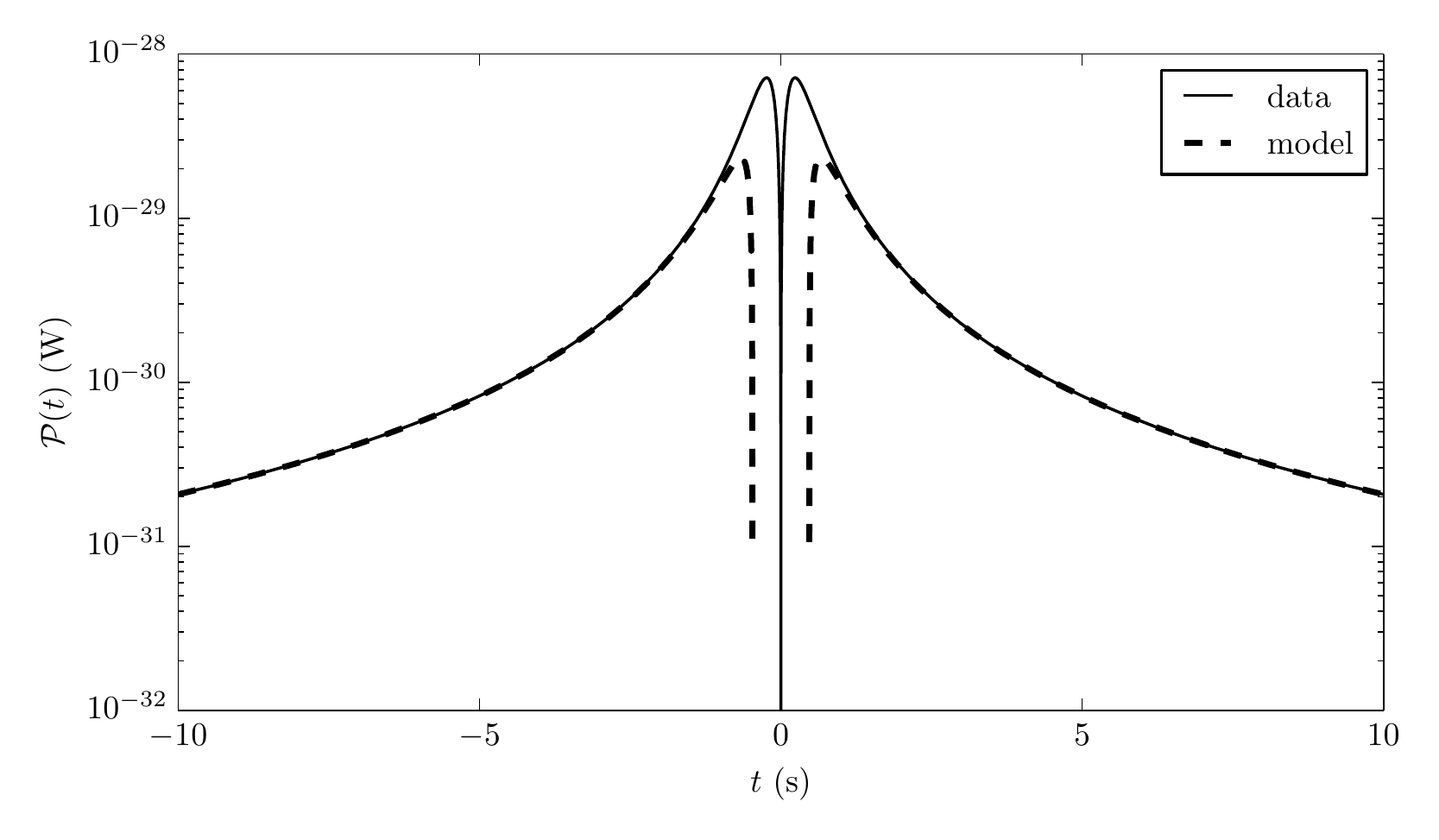}
	\caption{Example of a light curve for a speed $v = 0.001c$, impact parameter $b = 10^3\,$m, sail area $\mathcal{S} = 1\,$m$^2$, and P9 parameters $M = 5\,M_\oplus$ and $a^* = 0$ (solid line). The approximation of Eq.~(\ref{eq:DL}) leads to the dashed line.}
	\label{fig:light_curve}
\end{figure}

The expressions (\ref{eq:power}) and (\ref{eq:area}) (with ideal or perturbed geometrical quantities) allow us to compute the light curve received by SC0 as it passes by P9. A test result is shown in Fig.~\ref{fig:light_curve}. The main aspect of this test signal is that it is symmetrical, making the detection easier with respect to the background. Doppler effect would make it asymmetrical but due to the sub-relativistic speed it has negligible effects in our analysis. One can extract the parameters from the signal by using the following approximation, which is valid far from the minimal approach position $vt\gg b$
\begin{eqnarray}
\mathcal{P}(t) &= \frac{S(t)}{4\pi r(t)^2} \int_{0}^{+\infty} E \frac{\d^2 N}{\d t\d E} \nonumber\\
&\equiv \frac{S(t)}{4\pi r(t)^2} \mathcal{P}_0 \nonumber\\
&\approx \frac{\mathcal{P}_0\mathcal{S}}{4\pi}\frac{1}{(vt)^2}\left(1 - 2\left(\frac{b}{vt}\right)^2\right) \,, \label{eq:DL}
\end{eqnarray}
as can be seen in Fig.~\ref{fig:light_curve}. This approximation is valid in the straight line trajectory approximation, which is a good approximation as we will see below. In Fig.~\ref{fig:examples} we show the light curves for different setups as summarized in Table~\ref{tab:setup}. According to Eq.~(\ref{eq:DL}) one has to draw the detection signal with a rescaled time
\begin{equation}
t_0 = \left(\frac{3\times10^4\,{\rm m}}{b}\right)\,{\rm s}\,,
\end{equation}
in order to display all signals of Fig.~\ref{fig:examples} in the same plot. This is only in the favourable setup 1 that one gets an order of magnitude for the radio signal that is comparable with the currently most precise (Earth-based) detection tools. For example, the project Breakthrough Listen\footnote{\url{https://breakthroughinitiatives.org/initiative/1}} aims at detecting GHz signals from nearby stars to search for artificial signals as hints of advanced civilizations. Ref.~\cite{Sheikh2020} claims a minimal flux detection of $7.14\times10^{-26}\,$W$\cdot$m$^{-2}$ using the Green Bank Telescope -- a 100 meters diameter collecting antenna \cite{MacMahon_2018}. We do not expect the signal extraction from ambient noise to be any more difficult in P9 neighbourhood than on Earth. In Fig.~\ref{fig:examples} we show also the results with the exact trajectory calculations taking into account the gravitational well of P9. We see that for the considered setups the effect is very small.

\begin{figure}[t]
	\centering
	\includegraphics[scale = 1]{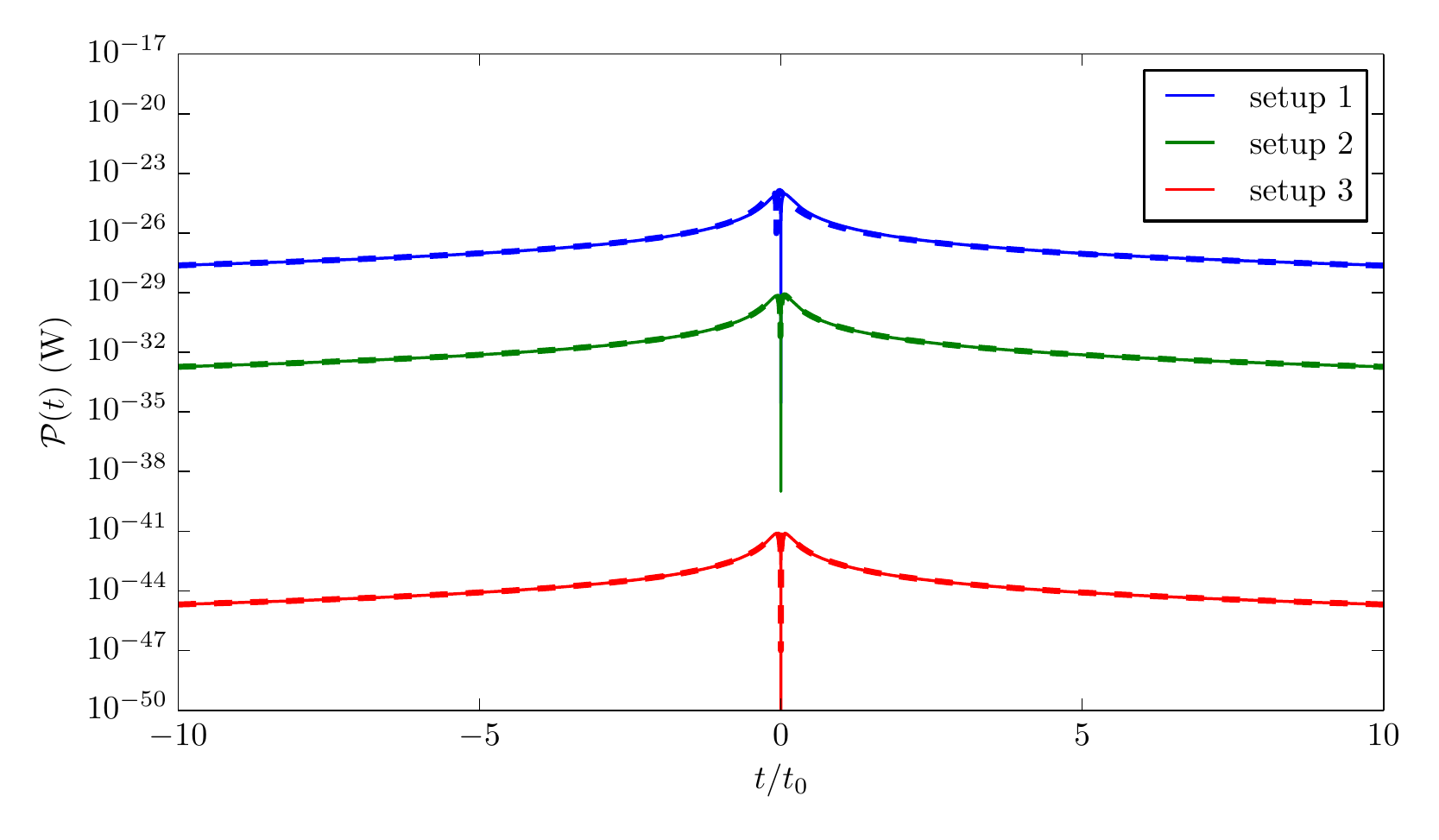}
	\caption{Radio signals received by SC0 for different setups with parameters given in Table~\ref{tab:setup}. We show both the ideal straight trajectories (plain lines) and the fully perturbed trajectories (dashed lines).}
	\label{fig:examples}
\end{figure}

\begin{table}
	\caption{\label{tab:setup}Parameters of the P9 and SC0 setups used in Fig.~\ref{fig:examples}.}
	\footnotesize
	\begin{tabular}{@{}lllll}
		\br
		setup & $M$ & $a^*$ & $b$ & $\mathcal{S}$ \\
		\mr
		setup 1 & $5\,M_\oplus$ & $0.99$ & $10^5\,$m & $100\,$m$^2$ \\
		setup 2 & $5\,M_\oplus$ & $0$ & $10^6\,$m & $10\,$m$^2$ \\
		setup 3 & $10\,M_\oplus$ & $0$ & $1\,$AU & $1\,$m$^2$\\
		\br
	\end{tabular}\\

\end{table}
\normalsize

P9 mass $M$ affects the energy of emission and thus its power. The resulting signal is proportional to the inverse of the mass squared (temperature squared). The degeneracy in mass is small for P9, hence we expect a $\mathcal{O}(10)$ factor at best when going from higher masses to lower masses as permitted by current constraints. P9 spin $a^*$ affects the emission rate and the power received, and we know that the signal can be enhanced by a factor of $\mathcal{O}(100)$ for photons when the spin is near extremal \cite{Page1976,Constraints_gamma_1}. The signal reception is proportional to the sail area $\mathcal{S}$, so multiplying the area by $\mathcal{O}(10)$ gives an amplification factor of $\mathcal{O}(100)$. The impact parameter $b$ fixes the minimum distance $r(t)$ that can be achieved, so the peak result is inversely proportional to $b^2$. The impact parameter on the other hand is a highly random parameter, which depends on the density of spacecrafts launched in the direction of the orbit of P9.

This fly-by scenario would be a totally independent and electromagnetic based mean of detection of P9 if it were a BH. However, as we have seen in this section, the CMB intensity is much larger than the Hawking radiation intensity in the considered energy range. In setup 1, the most favorable scenario considered here, this represents a signal-to-CMB ratio of $\sim 10^{-10}$, which is even smaller than the CMB fluctuations. Thus it seems unrealistic to detect P9 during a fly-by with this method. However, as we will see in the next section, the same method can be used to search for lighter BHs.

\section{Other perspectives}
\label{sec:others}

\subsection{Observation by a satellite}

\begin{figure}[t]
	\centering
	\includegraphics[scale = 1]{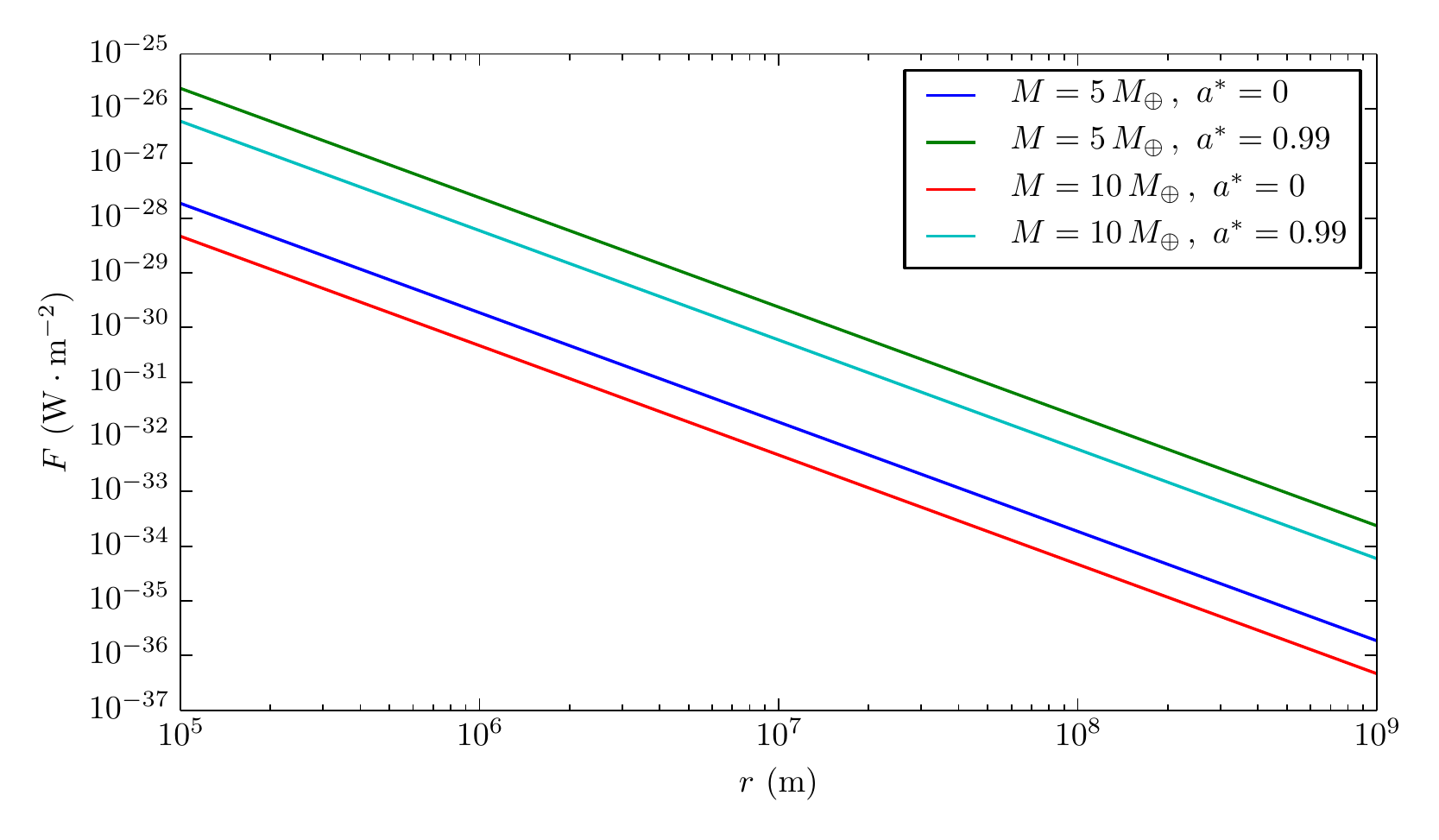}
	\caption{Radio flux as a function of orbit radius for different P9 masses $M = \{5,10\}M_\oplus$ and spins $a^* = \{0,0.99\}$.}
	\label{fig:satellite}
\end{figure}

%Finally, we point out that our proposal of Hawking radiation detection during a fly-by can be viewed as a complementary mean of detection of P9, would it be a BH.
Optimizations of proposals presented in \cite{P9_2020_1,P9_2020_2}, while taking into account the trajectory shifts estimated in \cite{P9_2020_3}, or proposal \cite{P9_2020_4}, may lead to a drastic reduction in the possible sky localization of P9 along its already constrained orbit. Therefore, with a more precise determination of its localization and if P9 still appears as a BH, it will be of utmost importance to send a mission orbiting P9, or at least to try to achieve the closest possible fly-by for a radio mission as described in this work. Hawking radiation would be the only direct measurement of the presence of P9, gravitational perturbations being only indirect evidence. The in situ measure of radio emission will give access to the form and properties of the BH horizon, thus giving exciting prospects for BH and fundamental physics. In case of satellization of a spacecraft around P9, Fig.~\ref{fig:satellite} shows the radio flux $F$ as a function of the orbit radius $r$, defined as
\begin{equation}
F = \frac{1}{4\pi r^2} \int_{0}^{+\infty} E \frac{\d^2 N}{\d t\d E} \d E\,.
\end{equation}
A satellized mission would offer the possibility to extensively study the electromagnetic emission of P9 if it were a BH. For example, one could imagine a directional parabolic antenna focused on the BH localization to reduce the impact of the CMB background, which will in addition be reduced by the screening due to the black hole horizon. This requires to reach high-precision focus of the order of the wavelength ($\sim\,$cm) at more than a hundred kilometers. It would then be achievable to distinguish the shadow of the BH on the CMB background, making it an indirect observation of P9. The long-exposure measure of this shadow as the spacecraft orbits P9 should be compared with the numerically predictable shadow on a constant CMB background, and then a radio signal coming out of the center of the shadow could be searched for. One can also imagine a mission constituted of two spacecrafts, one of which acting as a CMB shield screening the background and aligned with P9 and the antenna on the opposite location on the orbit. Then a signal received from P9 would constitute a direct measurement of its genuine emission.

\begin{figure}[t]
	\centering
	\includegraphics[scale = 1]{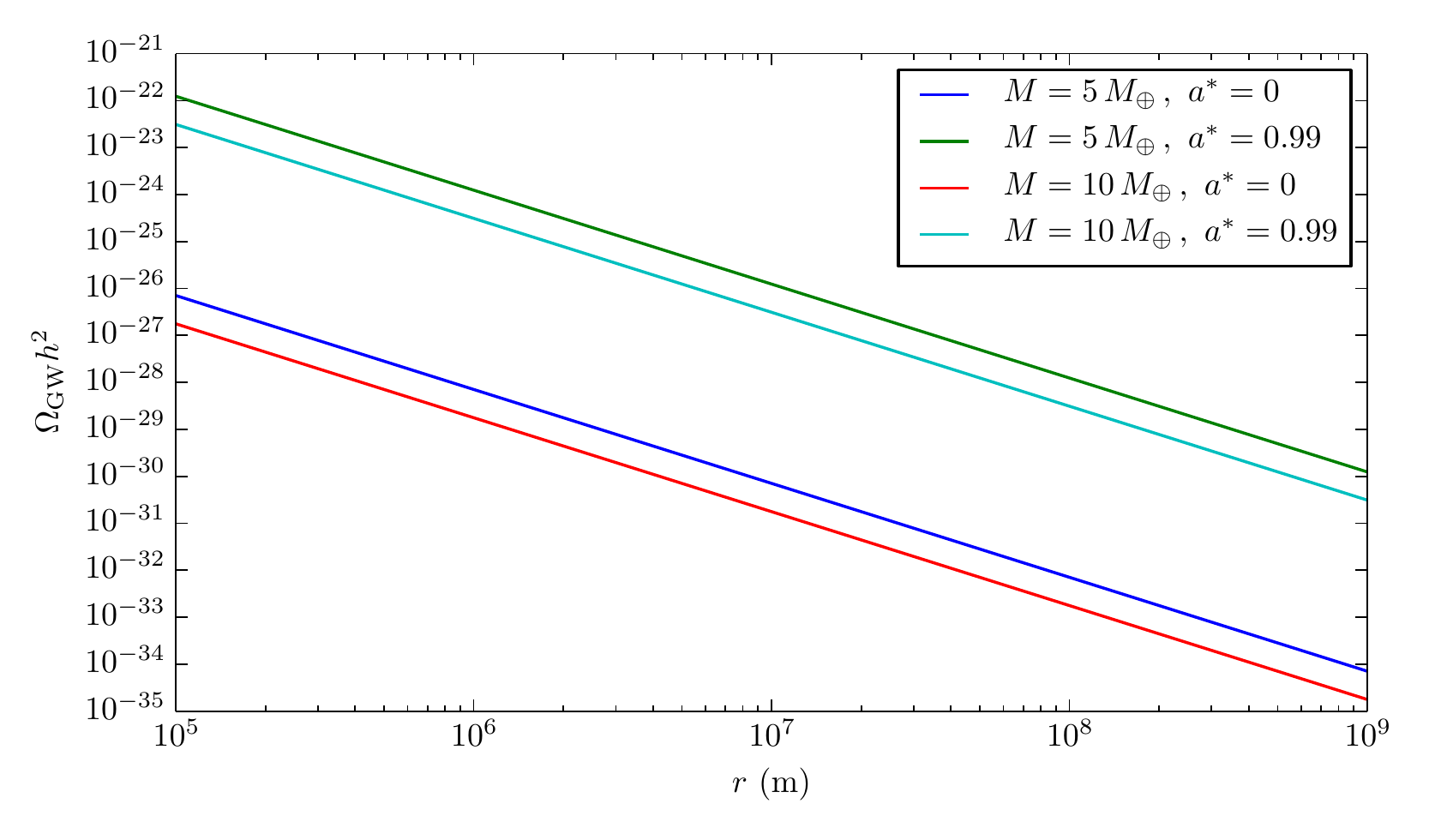}
	\caption{GWs density as a function of orbit radius for different P9 masses $M = \{5,10\}M_\oplus$ and spins $a^* = \{0,0.99\}$.}
	\label{fig:GW}
\end{figure}

Another direct probe of the presence of such a heavy BH via Hawking radiation is the emission of gravitational waves (GWs). In a semi-classical view of gravity, GWs are dual to massless spin 2 particles named gravitons. If the graviton is indeed a fundamental particle, it can be expected to be emitted by Hawking radiation. It has already been conjectured that graviton emission by PBH evaporation in the primordial universe could constitute a stochastic background carrying information on the first seconds after the Big Bang \cite{GW2009,GW2011,GW2016,GW2020,GW2020_2}. The detection of this high-frequency background remains a technical challenge. The amount of GWs emitted by present day BHs is again usually considered too low to be detectable \textit{from Earth}. If one were to put spacecrafts in orbit around P9, search for such gravitational waves would be of utmost importance to probe the existence and properties of the gravitons, constituting a portal to quantum gravity. In Fig.~\ref{fig:GW} we show the density of GHz GWs that such a satellite would receive as a function of its orbit radius
\begin{equation}
\Omega\mrm{GW} = \frac{1}{c\rho\mrm{c}}\left(\frac{H_0}{100\,{\rm km}\cdot{\rm s}^{-1}\cdot{\rm Mpc}^{-1}}\right)^2\frac{1}{4\pi r^2} \int_{0}^{+\infty} E \frac{\d^2 N}{\d t\d E} \d E\,,
\end{equation}
where $c$ is the speed of light, $\rho\mrm{c}\approx 8.523\times 10^{-30}\,$g$\cdot$cm$^{-3}$ is the critical density and $H_0 \equiv h\times (100\,{\rm km}\cdot{\rm s}^{-1}\cdot{\rm Mpc}^{-1})$ with $h \approx 0.67$ the reduced Hubble constant \cite{Planck2018}. Since it would constitute a constant signal, extraction from the noise may be easy. We see from Fig.~\ref{fig:GW} that a high P9 spin can increase the amplitude of GWs by 4 orders of magnitude \cite{Page1976,GW2016}. The signal-to-noise ratio is however complicated to forecast since no GW detector has for now explored the GHz domain. It is nevertheless probable that a lot of cataclysmic phenomena in the Universe may produce short GW bursts that fall into the GHz band.

Let us finally point out that all predictions made here depend on the validity of the calculation of Hawking radiation as it is predicted by our current knowledge. Any negative measurement of P9 emission would represent the first upper bound on the intensity of Hawking radiation, should this radiation behave differently from what we expect.

\subsection{Lighter PBHs}

\begin{figure}[t]
	\centering
	\includegraphics[scale = 1]{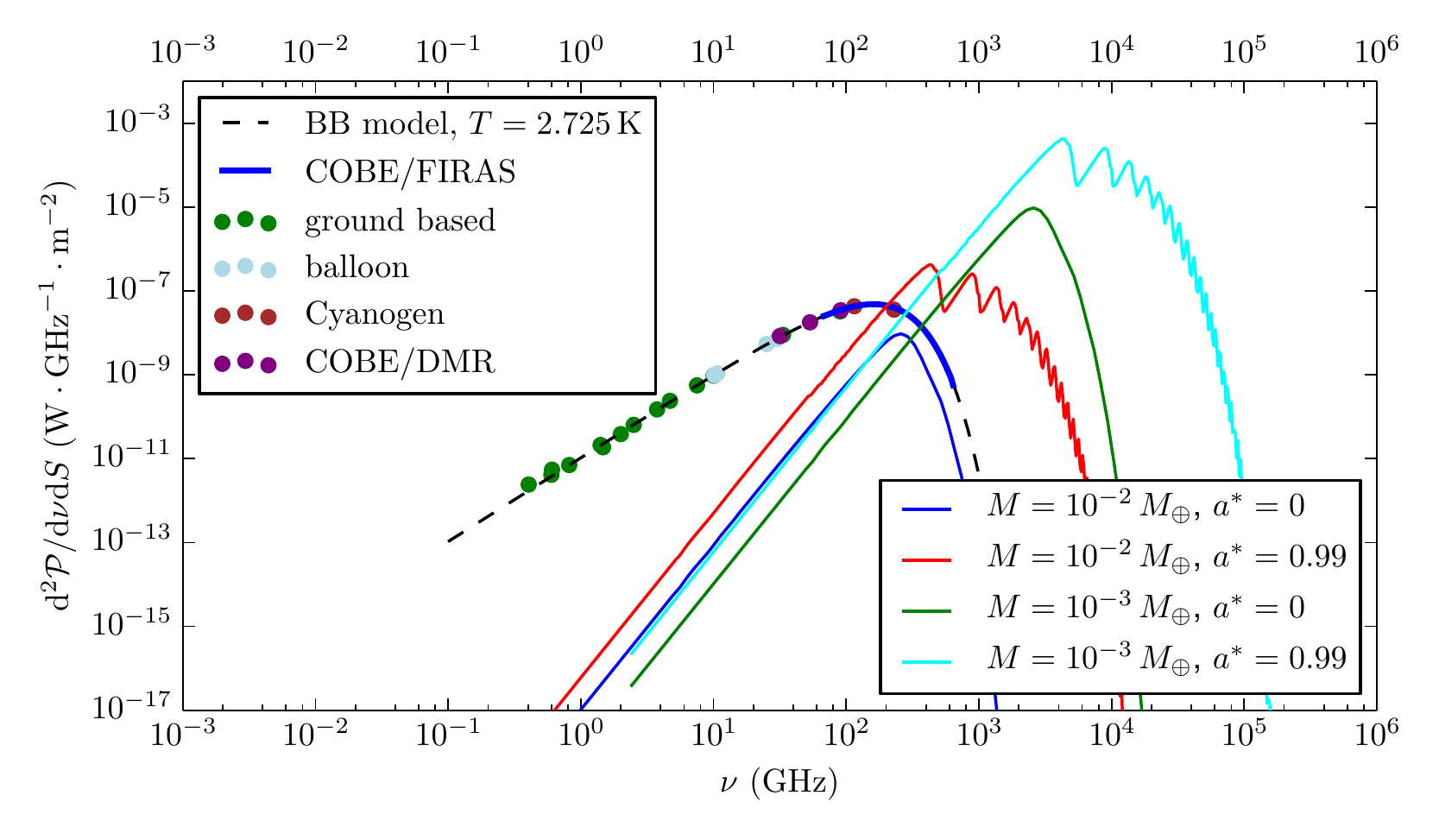}
	\vspace{-1cm}
	\caption{Total power emission of photons by light PBHs as a function of frequency for different masses $M = \{10^{-3},10^{-2}\}M_\oplus$ and $a^* = \{0,0.99\}$ (bottom solid lines). The CMB spectrum is the same as in Fig.~\ref{fig:band}.}
	\label{fig:mercury}
\end{figure}

We can extrapolate the present discussion to lighter BHs that could have been captured by the gravitational field of the Sun \cite{Schneider2020}. A $\sim 0.01\,M_\oplus$ body of the mass of Mercury would emit in the same band of energy and with an intensity close to the CMB, because its temperature is $T\sim T\mrm{CMB}$, making its putative detectability much easier. In Fig.~\ref{fig:mercury} we show the same emissivity as in Fig.~\ref{fig:band} but for lighter PBHs. We can see that the energy range of the emission lies in the CMB peak at $\sim 2.7\,$K, but the emissivity is higher than the CMB one for PBHs with mass $M\lesssim 10^{-2}\,M_\oplus$. There is no evidence of such light hidden bodies in the outer Solar system, contrarily to the P9 gravitational perturbations, but the expected perturbations would be too small to be detected by TNOs orbits clustering. During the last years, several light bodies have been discovered in the outer Kuiper belt: Eris, Haumea, MakeMake... showing that a large population of such objects can exist beyond Neptune's orbit. However, the fraction of DM that these objects can represent is more tightly constrained by microlensing than terrestrial mass objects, with a fraction $f\lesssim 1\%$ \cite{Carr2020}. A satellized mission around one of those objetcs as described in the previous paragraph may lead to the first direct measurement of Hawking radiation, thus allowing comparisons with models alternative to the classical Hawking prediction.

\section{Conclusion}

In this exploratory work we have proposed a new way to probe the presence of a hypothetical Planet 9 in the outer Solar System if it were actually a black hole, by using a Breakthrough Starshot-like fleet of nano-spacecrafts. Considering the difficulties of measuring tiny longitudinal or transverse displacements that P9 would induce on a spacecraft during a pass-by, mostly related to the fact that trajectory perturbations arising from the interstellar medium would be of the same order, we propose to measure \textit{in situ} the Hawking radiation emitted by P9 in the form of GHz radio photons. This method has two main advantages, first it is not affected by the trajectory noise because it only relies on classical on-board electromagnetic detection, second it would be a unique occasion to measure and thus prove the existence of Hawking radiation, a long-standing prediction of black hole thermodynamics. The principal difficulty is to measure a very faint signal in the radio GHz band, with an amplitude inversely proportional to the square of the impact parameter $b$, therefore requiring either great luck or a multitude of spacecrafts in order to reach a fly-by of P9 at $\sim100\,$km distance, or the use of an extremely precise radio detection technology. Furthermore it seems unrealistic to extract this faint signal from the dominant CMB contribution during a fly-by mission. Nevertheless, if P9 were indirectly localized using for example spacecraft trajectory measurements or LSST flares, an orbital mission would be of great importance to study the properties of black holes and Hawking radiation, and would allow for more advanced measurement techniques, for example by screening the CMB and focusing the antenna on P9. Finally, we extrapolated the present discussion to lighter PBHs that could have been captured by the gravitational field of the Sun and concluded on the advantages of these configurations for the detection of Hawking radiation, since their emissivity would be dominant as compared to the CMB.

\section*{References}

\bibliographystyle{unsrt}
\bibliography{biblio}

\end{document}